\documentclass{article}

\usepackage[english]{babel}

\usepackage[letterpaper,top=2cm,bottom=2cm,left=3cm,right=3cm,marginparwidth=1.75cm]{geometry}

\usepackage{amsmath}
\usepackage{amssymb}
\usepackage{mathpazo}
\usepackage[mathpazo]{flexisym}
\usepackage{breqn}
\usepackage{graphicx}
\usepackage[colorlinks=true, allcolors=blue]{hyperref}
\usepackage{bm}
\usepackage{amsthm}
\theoremstyle{definition}

\usepackage{soul}

\usepackage{tikz-cd}

\usepackage{authblk}
\title{Coordination without communication: beyond optimisation and geometric Brownian motion}
\author{G J Milburn,\vskip 0.1 truecm
School of Mathematics and Physics, University of Sussex, Brighton, BN1 9RH, UK.\vskip 0.1 truecm
 and National Centre for Quantum Computing, Rutherford Appleton Laboratory, Harwell Campus, Didcot, Oxfordshire, OX11 0QX UK\\
A K Ringsmuth, \vskip 0.1 truecm
Wegener Center for Climate and Global Change, University of Graz, Graz, Austria \vskip 0.1 truecm
and Complexity Science Hub, Vienna, Austria.
}

\begin{document}
\maketitle

\begin{abstract}
We introduce a physically grounded framework for coordination in a population based on information-constrained feedback in a partially observed stochastic dynamical system. Population size evolves as a continuous-time birth–death Markov process whose transition rates respond to a shared stochastic measurement signal correlated with the underlying population state. Individuals neither communicate directly nor optimise strategies; instead, coordination emerges from macro-to-micro feedback mediated by imperfect common information. We show that geometric Brownian motion arises as a limiting case of the conditional dynamics when measurement strength and population statistics satisfy suitable conditions. More generally, varying the signal-to-noise properties of the measurement channel produces a wider class of stochastic growth processes, including diffusive and jump-like regimes, even though ensemble-average growth remains exponential. In an appropriate limit the framework recovers the stochastic multiplicative growth model of Peters and Adamou, providing a physical interpretation of coordination as inference and feedback under partial observability.

\end{abstract}

\section{Introduction}
A recurrent motif in complex systems is feedback across levels of organisation, whereby \textcolor{black}{local(micro)} entities react to  \textcolor{black}{global(macro)} \textcolor{black}{extensive variables}. Such macro-to-micro feedback\st{s} is well known to generate large-scale order in physical, biological, and social systems without central control\cite{RevModPhys.81.591}. In biological and social contexts, these emergent regularities are often described as coordination or cooperation. Coordination refers to the alignment of states or behaviours arising from coupling and feedback, and need not involve conflict between individual and collective objectives. Cooperation, in the strong sense used in evolutionary game theory, refers to individually costly actions that generate collective benefits in settings where defection would be advantageous to the individual. However, “cooperation” is also often used more loosely to denote coordinated behaviour; where the distinction matters, we make it explicit.

Large-scale social and ecological crises such as pandemics and climate change are often conceptualised as failures of cooperation. Even when all relevant actors agree that collective harm should be avoided, locally reinforced incentives, spatial separation of costs and benefits, and weak or indirect feedback can sustain dynamics that drive systems towards tragedy. More generally, large-scale collective failures can arise when local actions couple only weakly and indirectly to emergent system-level states. Under such conditions, collective outcomes are shaped less by individual objectives than by the mechanisms of macro-to-micro feedback. In multiscale systems, therefore, persistent failure can arise from coordination problems without requiring pervasive antagonism or defection from agreements or norms.

Most formal models of coordination and cooperation invoke externally specified objectives. In game-theoretic formulations, behaviour is organised around payoff functions, typically interpreted as representations of individual utility, as in canonical social dilemma models such as the prisoner’s dilemma \cite{Smith-Price}. These make nontrivial assumptions about agents’ internal states; at least, agents recognise they are playing a game and know its rules. Optimisation-based approaches rely on abstract cost functions, as in ant colony optimisation \cite{ACO} with relatively simple agents, or learning models with sophisticated representations of agents’ internal states \cite{RL-optimisation}. In economics, markets are commonly treated as decentralised coordination mechanisms, in which prices summarise system-level information and guide local behaviour, an interpretation justified by the claim that market equilibria correspond to solutions of an implicit global optimisation problem under the restrictive assumptions of mainstream economic theory. It is also assumed that underlying macroscopic  \textcolor{black}{states} (e.g. relative scarcities) are stably and meaningfully encoded in price signals available to all agents. In all of these cases, the objective function is typically introduced by the modeller for analytical or computational convenience rather than derived from biophysical reality and, when it encodes the phenomenon to be explained, explanatory circularity results. 

These approaches also typically assume that all agents are well informed about the system-level variables relevant for coordination, such as prices or institutional rules. In many systems, however, the emergent macroscopic state relevant for coordinating micro constituents is only weakly observable, so that no single signal reliably couples local actions to global outcomes. This raises a fundamental question: can such phenomena be most effectively explained by adapting the above standard approaches, or might it instead be possible to use well-understood physical principles grounded in nonequilibrium thermodynamics without attributing any notion of agency or explicit optimisation to the system’s constituents? Would a successful answer along these lines be sufficient, or would there remain an explanatory gap?

A canonical example of a purely dynamical explanation for collective behaviour is the Vicsek model \cite{vicsek1995novel}, which shows how local alignment interactions combined with noise generate large-scale collective motion in systems of distinguishable individuals. Despite lacking planning, memory, or explicit optimisation, the model exhibits a phase transition from disordered motion at high noise or low density to ordered collective motion at low noise or high density.

Related perspectives demonstrate that cooperation-like behaviour can arise directly from physical interactions. Benjamin and Sarkar \cite{benjamin2020} mapped two-player social dilemma games onto a one-dimensional Ising model, showing how outcomes resembling cooperation emerge in the thermodynamic limit without intentional decision-making. In chemical and biochemical systems, autocatalytic networks provide an even starker illustration: Kauffman \cite{Kauffman1993OriginsOfOrder} characterised cooperation as the closure of catalytic interactions, where apparent goal-directedness emerges from network topology and feedback rather than optimisation or strategy. From this perspective, game-theoretic descriptions operate at a higher level of abstraction. Payoff matrices and strategies can be understood as compressed representations of the consequences of underlying interaction dynamics, rather than as primitive drivers of behaviour. Such descriptions become relevant only once agents possess cognitive or institutional structures that allow them to internalise and act upon these consequences.

More recently, stochastic growth processes have been proposed as a basis for understanding how cooperation emerges. Peters and Adamou \cite{Peters} modelled a population of individuals, each with a resource stock evolving under geometric Brownian motion (GBM). They found that repeated pooling and sharing of resources reduced the volatility experienced by individuals, thereby increasing their time-averaged growth rates. Fant et al. extended this approach using stochastic differential equations, with GBM as a baseline process \cite{PhysRevE.108.L012401}. Building on the same stochastic foundation but incorporating explicit network structure to model pooling and sharing, Stojkoski et al. \cite{PhysRevE.99.062312} demonstrated that interactions between stochastic growth and network topology can produce large discrepancies in observed behaviour. In these models, what is labelled cooperation is more precisely understood as coordination: alignment that improves long-term viability without a fundamental conflict between individual and collective outcomes. GBM also underlies the Black–Scholes model for option pricing \cite{Black-Scholes}, whose empirical limitations have motivated various generalisations, including jump–diffusion models \cite{Haugh2010BeyondBlackScholes,e22121432,BurgerKliaras2013JumpDiffusion}.

Despite their diversity, the models reviewed above share a common structural feature: collective effects are implemented through explicitly specified coupling rules whose form is known a priori. The quantities that characterise macroscopic observables relevant to collective behaviour, such as alignment fields, catalytic closures, effective growth rates, or pooling and sharing protocols, are built directly into the dynamics rather than inferred by agents from observations. In some cases these couplings can be interpreted as macro-to-micro feedback, while in others they act purely through statistical aggregation, but in all cases the relevant collective variables are prespecified. By contrast, many biological and social systems operate under partial observability: macroscopic variables that regulate behaviour, such as population size, density, or aggregate activity, are not directly accessible to individual constituents, and feedback is mediated by noisy, indirect signals from which global properties must be estimated.

In this work, we address this gap using a framework inspired by models of feedback in partially observed Markov processes. We model a population of agents evolving according to an underlying continuous-time Markov process, whose transition rates are modulated by a partially observed stochastic signal corresponding to a global measurement of the population state. This signal constitutes an effective stochastic continuous measurement of an emergent macroscopic variable. Agents may estimate this variable and infer parameters of the underlying dynamics from observation histories, subject to an uncertainty constraint, which we calculate. By making inference an explicit component of macro-to-micro feedback, our framework extends existing dynamical accounts of coordination and provides a physically grounded bridge to standard higher-level descriptions of cooperation.

\section{Model}

Let $n_\alpha(t)>0$ be the number of agents of type $\alpha$ and denote these by a  $D$ dimensional column vector $ {\bm n}$ with components $(n_1,n_2,\ldots n_D)$. There is no upper constraint on the number of individuals of each type. Let ${\bm e}_\alpha$ be the unit vector for type-$\alpha$. The two elementary events:
\begin{itemize}
    \item increase $ {\bm n}\rightarrow  {\bm n}+{\bm e}_\alpha$ 
    \item decrease  $ {\bm n}\rightarrow  {\bm n}-{\bm e}_\alpha$  when $n_\alpha >0$
\end{itemize}
Each `jump' has an arbitrary non negative rate functional $w_\alpha^{+}({\bm n})$ or $w_\alpha^{-}({\bm n})$ that may depend on the full composition ${\bm n}$. The probability distribution for the number of agents of each type obeys a birth-death Markov process given by
\begin{equation}
\frac{\partial}{\partial t}P({\bm n},t)=\sum_{\alpha=1}^{N}
\Big[ w_\alpha^{+}({\bm n}-{\bm e}_\alpha)\,P({\bm n}-{\bm e}_\alpha,t)+ w_\alpha^{-}({\bm n}+{\bm e}_\alpha)\, P({\bm n}+{\bm e}_\alpha,t)- \big(w_\alpha^{+}({\bm n})+w_\alpha^{-}({\bm n})\big)\,P({\bm n},t) \Big ],
\end{equation}
with the convention that terms involving negative components are dropped. The boundary conditions are chosen so that the mean $\bar{n}_\alpha$ number of each type of agent is a finite positive real number for all time and $p_{\bm n}(t)=0$ for $n_\alpha<0$.  The first two terms are the gain of probability into ${\bm n}$ from neighbouring states that jump into ${\bm n}$. The last term is the loss of probability from ${\bm n}$ due to all possible $\pm1$ jumps out of $\mathbf{n}$. We refer to this as the {\em unconditional} master equation. We will contrast it with a {\em conditional} master equation where a partially observed process monitors the population of each type. 

The variables ${\bm n}(t)$ are unknown; they are 'hidden' from each agent. What agents do have access to is partially observed stochastic process ${\bm y}(t)$. Given the observations, the agent can predict the value for $n_\alpha(t)$, which we take to be equivalent to estimating the conditional distribution $p_{{\bm n},c}(t)$,  of the state given the observation history up to time  $t$. Based on the estimation, the agent computes the conditional means $\bar{\bm{n}}_{c}(t)$. The {\em innovation process} is defined as the difference between the actual observation, $d\bm{y}$ and the predicted observation, $\chi\bar{\bm{n}}_{c,t}dt $, where $\chi>0$, based on the conditional state at time $t$. In our case the innovation is a white noise (Brownian) process, $d\bm{y}-\chi\bar{\bm{n}}_{c,t} dt =\sqrt{\kappa}d\bm{W}$. In the appendix we construct a model by taking a continuous-time limit of a sequence of discrete-time uncertain measurements of the population number.

The evolution of the posterior probability  $p_{{\bm n},c}(t)$ is governed by a Kushner-Stratonovich-type equation or a filtering master equation, which couples the state transition dynamics (from the original master equation), and the observation update. The equation of motion for $p_{{\bm n},c}(t)$ we call the {\em conditional } master equation.  The conditional master equation  can take various forms, depending on the nature of the observations,  and is non linear in the conditional probability. This is a reflection of the fact that the future observations must be consistent with past observations.  For each population type, there is a corresponding measurement record ${\bm y}(t) = (y_1(t), \dots, y_D(t))$. The observations are given by an Ito stochastic differential equation 
\begin{equation}
\label{general-current}
 d{\bm y}(t) = \chi \bar{{\bm n}}_c(t)dt  + d{\bm W}(t) \equiv  {\bm I}(t)dt 
\end{equation}
 $d{\bm W}$ Weiner process increment (Gaussian noise) with covariance matrix  $\Sigma \in \mathbb{R}^{D \times D}$.  Averaging over the noise in Eq. (\ref{general-current}) we see that 
\begin{equation}
\label{average-current}
    \bar{{\bm I}}=\frac{d\bar{\bm{y}}}{dt}=\chi \bar{{\bm n}}(t)
\end{equation}
This suggests that $ {\bm I}(t)$ can be regarded as a {\em measurement current}. While the actual population ${\bm n}(t)$ is unknown, the measurement currents $ {\bm I}(t)$ are known globally. They are 'broadcast' to every individual in the system.  The integral of the measurement current up to time $t$ gives the measurement record, $y(t)$, at that time. We will refer to the integrated stochastic process, $y(t)$ as the {\em measurement signal}. 

A biological interpretation of the previous definitions may help understand the relationship between population, measurement current and measurement signal.  Suppose that there is only a single kind of agent -- a single phenotype -- and that each agent produces a metabolic product that becomes available to all to access. The total amount of this product at any time is the measurement signal $y(t)$. The rate of production is the measurement current, $I(t)$. The only way agents are aware of other agents is by direct knowledge of $I(t)$. We now define a measure of coordination, for this kind of agent, as the rate of production $I(t)$. More coordinated communities have a higher rate of production. The measurement signal, $y(t)$, is the cumulative amount of this chemical in the community or the accumulated wealth.  

We introduce coordination as feedback.  In terms of the previous biological interpretation, suppose that the rate of production increases linearly with the number of agents due to coordination.  Agents outside the community detect the rate of product growth and find it more attractive to join communities that have a high rate of production, as they are more coordinated. This leads to a positive feedback process. Why might agents evolve to focus on the rate of production rather than on net wealth? In a biological setting, life is uncertain and agents need to accumulate wealth as fast as possible. What matters is not the acquired wealth of the community at any time, but the rate of wealth creation at the time an agent joins.  

We will assume instantaneous feedback. That is to say, there is no time delay between the measurement current $ {\bm I}(t)$ and how each agent responds to it. Roughly speaking, it is as if each agent has immediate access to the collective measurement record and then adjusts their transition rates  accordingly.  The feedback vector values change the transition rates $w_\alpha^{\pm}(\mathbf{n})\rightarrow w_\alpha^{\pm}({\bm n},  {\bm I} )$ making them time dependent.  

\section{Example: single phenotype. }

We will illustrate the description using a very simple model with only one kind of agent. At any time the number of such agents is $n(t)$ but this value is unknown but is certainly non negative. We specify the population in terms of a probability $p_n(t)$ to find $n$ agents at time $t$.  We refer to this as the {\em unconditional probability distribution}. It obeys the Markov master equation
\begin{equation}
\label{master-equation}
    \frac{dp_n(t)}{dt}=t^+(n-1)p_{n-1}(t)+t^{-}(n+1)p_{n+1}(t)-[t^+(n)+t^-(n)]p_n(t)
\end{equation}
The two point processes are 
\begin{eqnarray}
    n \rightarrow n+1 &&\ \ \ \ \mbox{rate}\ \ \ t^{+}(n)\\
    n \rightarrow n-1 &&\ \ \ \ \mbox{rate}\ \ \ t^{-}(n)
\end{eqnarray}
The boundary conditions are $p_n(t) =0 , n< 0$, $\sum_{n} p_{n}(t) = 1 \ \forall \ t$ and $ p_n(t) \rightarrow  0\ \mbox{as  }  n \rightarrow \infty$ such the the average number, $\bar{n}$ is bounded.   The corresponding rate equations are the rate of change of the unconditional mean number, 
\begin{equation}
    \frac{d\bar{n}}{dt}= \overline{t^+(n)}- \overline{t^-(n)}
\end{equation}
where 
\begin{equation}
    \overline {f(n)}= \sum_{n=0}^\infty p_n(t) f(n)
\end{equation}

The measurement current is given by 
\begin{equation}
    I(t)=\frac{dy}{dt}
\end{equation} 
The stochastic process for the measurement signal is 
\begin{equation}
    dy=I(t)dt = \chi \bar{n}_c(t) dt+\sqrt{\kappa}dW(t)
\end{equation}
where $dW$ is a Wiener increment,  $\bar{n}_c(t)$ is the conditional mean population conditioned on the entire stochastic history of $I(t)$ up to time $t$, $\chi>0$ is a rate that determines the instantaneous value of the measurement signal and $\kappa$ sets the size of the diffusive noise (see Appendix). The conditional mean is defined by
\begin{equation}
    \bar{n}_c(t)=\sum_{n=0}^\infty n p_{n,c}(t)
\end{equation}
and is non negative. We can think of this as the estimated, or predicted, value of $n(t)$ given the previous history of the observations. The innovation process  is defined as the difference between the actual observation and the predicted observation, in time $dt$, based on the prior state of the model. Here the innovation process is $dy-\chi\bar{n}_c dt=\sqrt{\kappa}dW$

As an example, let agents  join the space with probability $\gamma dt$, independent of any macroscopic signal, in a increment of time $dt$ and never leave.    The  master equation is
\begin{equation}
\label{pop-markov}
    \frac{dp_n}{dt} = \gamma (p_{n-1}-p_n)
\end{equation}
where $\gamma$ is the rate at which individuals are 'injected' into the system.  Assuming $p_n(0)=\delta_{n,0}$ the solution is 
\begin{equation}
    p_n(t) = \frac{(\gamma t)^n}{n!} e^{-\gamma t}
\end{equation}
The mean number and variance  are $\bar{n}(t)=\gamma t={\mathbb V}[n]$, a Poisson process.

\begin{figure}[h!]
    \centering
    \includegraphics[scale=0.4]{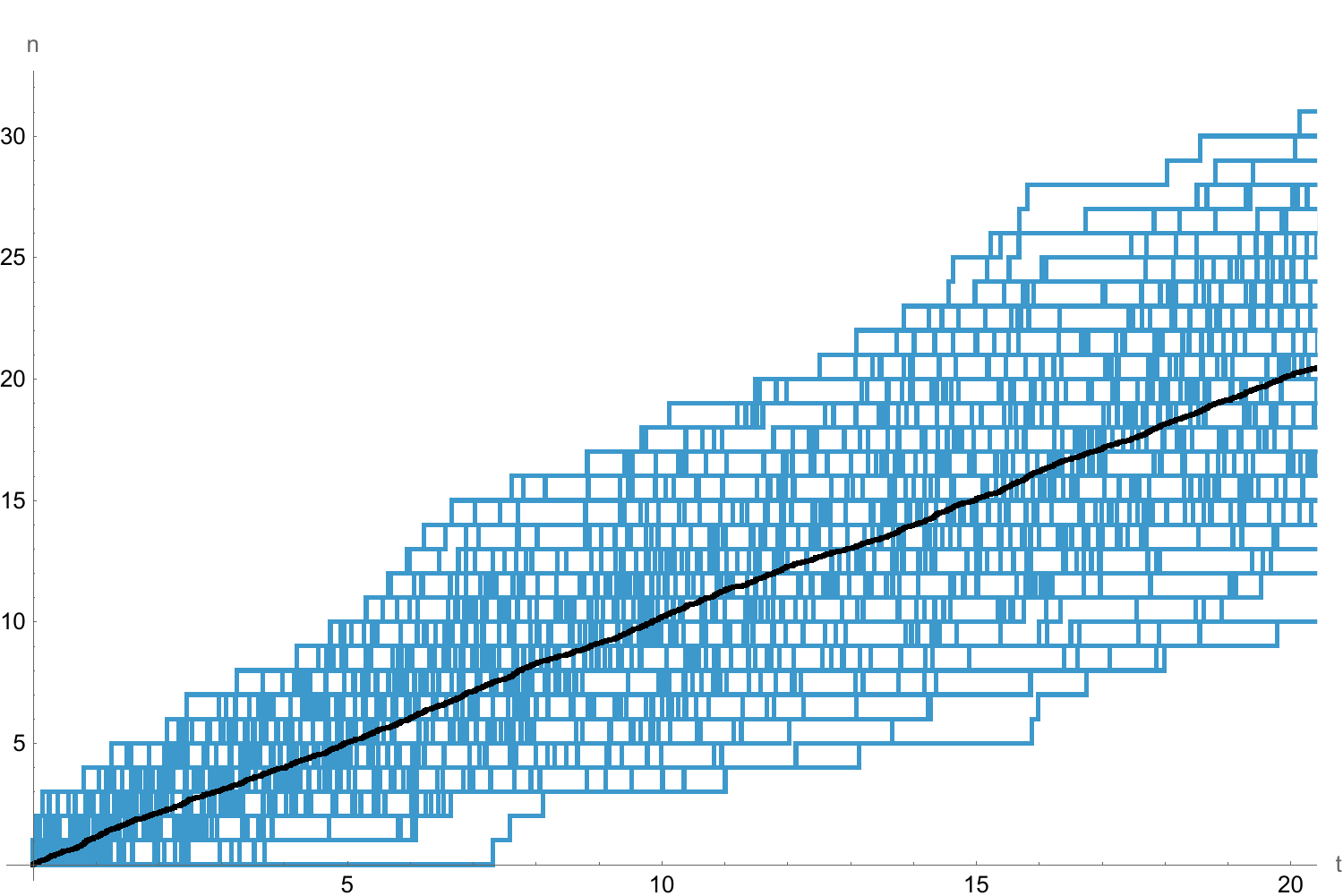}
    \caption{A sample of $100$  realisations of a stochastic population process modelled by Eq. (\ref{pop-markov}), with $\gamma=1.0$. Also shown is the empirical mean which approaches a line with slope $\gamma$}. 
    \label{trajectory-empty}
\end{figure}

The actual population at any time is a stochastic process with unit increments at random times, Fig(\ref{trajectory-empty}).
The key assumption here is that individuals enter the feature space independently of the decisions made by any other agent already in the feature space. No agent is `aware' of any other agent. 
Central to our treatment is the distinction between the {\em conditional} mean population, $\bar{n}_c(t)$ and the {\em unconditional} mean population $\bar{n}(t)$. The actual population in a sample at any time is unknown. A stochastic 'measurement signal' , $I(t)$, is broadcast to all agents in the population and depends, non deterministically, on the unknown population at any time. 

Starting with an initial condition of zero population, the measurement signal begins to change when the first agent enters the population. This event occurs at a random time. As agents continue to enter the feature space, the measurement signal changes. In the course of time we generate an entire history for the measurement signal as a stochastic function of time $I(t)$,
\begin{equation}
\label{measurement-current}
   I(t)dt= \chi \bar{n}_c(t) dt+\sqrt{\kappa}dW(t)\ ,
\end{equation}
where $dW$ is a Wiener increment,  $\bar{n}_c(t)>0$ is the {\em conditional mean population} conditioned on the entire stochastic history of $I(t)$ up to time $t$, $\chi>0$ is a rate that determines the instantaneous value of the measurement signal and $\kappa$ sets the size of the diffusive noise. If we ran that process again, from the same initial condition we would get a different record for the measurement signal. In each trial the current is conditional on the unknown population at any time $n(t)$. If we average over many trials, all starting from the same initial condition,  we get the unconditional mean $\bar{I}(t)$,
\begin{equation}
\label{mean-measurement-current}
\bar{I}(t)= \chi \bar{n}(t)\ ,
\end{equation}
The unconditional mean current is proportional to the unconditional mean of the population.

The added noise in Eq. (\ref{measurement-current}) could dominate the 'signal' deriving from the population. It would be useful to have a measure of the quality of the signal. This is the signal-to-noise ratio (SNR). The SNR is defined by first adding a response rate, $\tau^{-1},$ to the bare measurement current, Eq. (\ref{measurement-current})
\begin{equation}
\label{filter-measurement-current}
   dy= - \frac{1}{\tau}y dt+ \chi \bar{n}_c(t) dt+\sqrt{\kappa}dW(t)
\end{equation}
Integrating, we see that
\begin{equation}
y(T)= y(0)e^{- t/\tau}+\chi \int_0^T dt' e^{-(t-t')/\tau}  \bar{n}_c(t')+\sqrt{\kappa}\int_0^T dt' e^{- (t-t')/\tau}  dW(t')
\end{equation}
Let us take $y(0)=0$. The ensemble average signal is determined by the unconditional average, 
\begin{equation}
 \bar{y}(T)  = \chi \int_0^T dt' e^{- (t-t')/\tau}  \bar{n}(t')
\end{equation}
while the variance is given by 
\begin{equation}
     {\mathbb V}[y]= \frac{\kappa\tau}{2}(1-e^{-2 T/\tau})
\end{equation}
and we define the square of the signal to noise ratio as 
\begin{equation}
    SNR^2= \frac{(\bar{y}(T))^2}{{\mathbb V}[y(T)]}
\end{equation}
 If the integration time, $T$, is short compared to the response time, $\tau$, of the detector, $ T \ll \tau$, 
\begin{eqnarray}
\bar{y}(T) & \approx & \chi T \bar{n}(T) \\
    {\mathbb V}[y(T)]& \approx & \kappa T
\end{eqnarray} 
and 
\begin{equation}
\label{SNR}
    SNR^2=\frac{\chi^2 T}{\kappa} \bar{n}=\Gamma T\bar{n}
\end{equation}
where
\begin{equation}
    \Gamma =\frac{\chi^2}{\kappa}\ .
\end{equation}
If the noise dominates, the SNR is less than unity.  Intuitively, the SNR is a measure of the quality of the information in the measurement current when we integrate for a time $T$. The sensitivity of the measurement is defined as $SNR/\sqrt{T}$. The underlying markov process defining the fluctuations in the population is determined by the rate $\gamma$. We see that a `strong' measurement requires $\Gamma \gg\gamma $. In this limit, the signal can track changes in the population very easily for small integration times.  

We can write the stochastic measurement current directly in terms of the measurement strength $\Gamma$ by scaling the current by $\chi$, $y\rightarrow y'=y/\chi$, 
\begin{equation}
\label{scaled-filter-measurement-current}
dy'=  \bar{n}_c(t) dt+\frac{1}{\sqrt{\Gamma}}dW(t)
\end{equation}

In the appendix we give a detailed measurement model to define the conditional Markov process using Bayes' theorem. It is given by
\begin{equation}
\label{SME}
   dp_{c,n}= \gamma (p_{c,n-1}-p_{c,n})dt   +\sqrt{\Gamma}(n-\bar{n}_c)p_{c,n}dW(t)
\end{equation}
In this paper  the inital condition is fixed $p_n(0)=\delta_{n,0}$, the population starts at zero. The constant background rate ensures that it will never stay zero. 
The conditional mean then obeys the stochastic differential equation.
\begin{equation}
\label{conditional-mean}
    d\bar{n}_c(t) = \gamma dt +\sqrt{\Gamma}{\mathbb V}_c[n]dW
\end{equation}
where the conditional variance is defined by 
\begin{equation}
   {\mathbb V}_c[n] =\sum_{n=0}^\infty (n-\bar{n}_c)^2 p_{c,n}(t) 
\end{equation}



\begin{figure}
    \centering
    \includegraphics[scale=1.0]{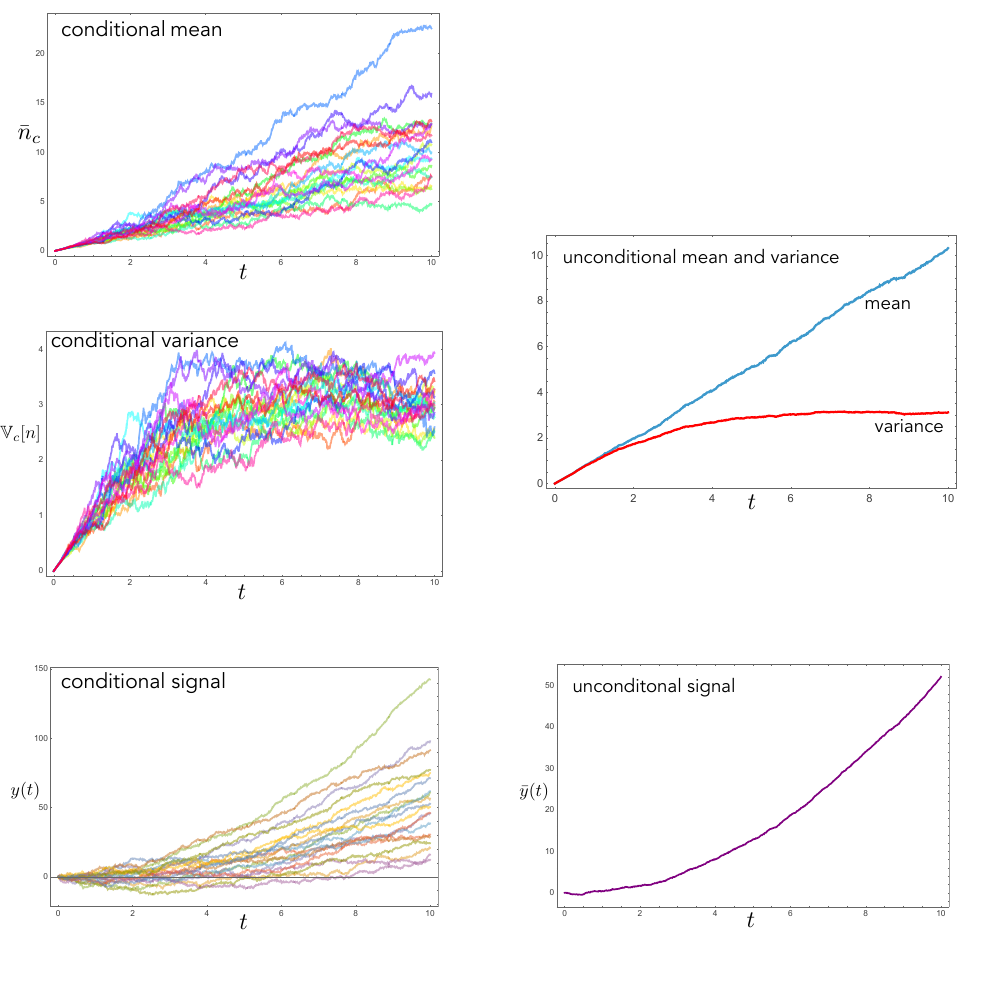}
    \caption{Weak measurement regime. Typical simulations of the conditional and unconditional mean population, conditional variance,  and the stochastic measurement signal as a function of time. Also shown is the ensemble average stochastic measurement signal $\bar{y}(t)$. Parameters are $\gamma= 1.0, \Gamma =0.1$.  }
    \label{no-feedback-weak}
\end{figure}

\begin{figure}
    \centering
    \includegraphics[scale=1.0]{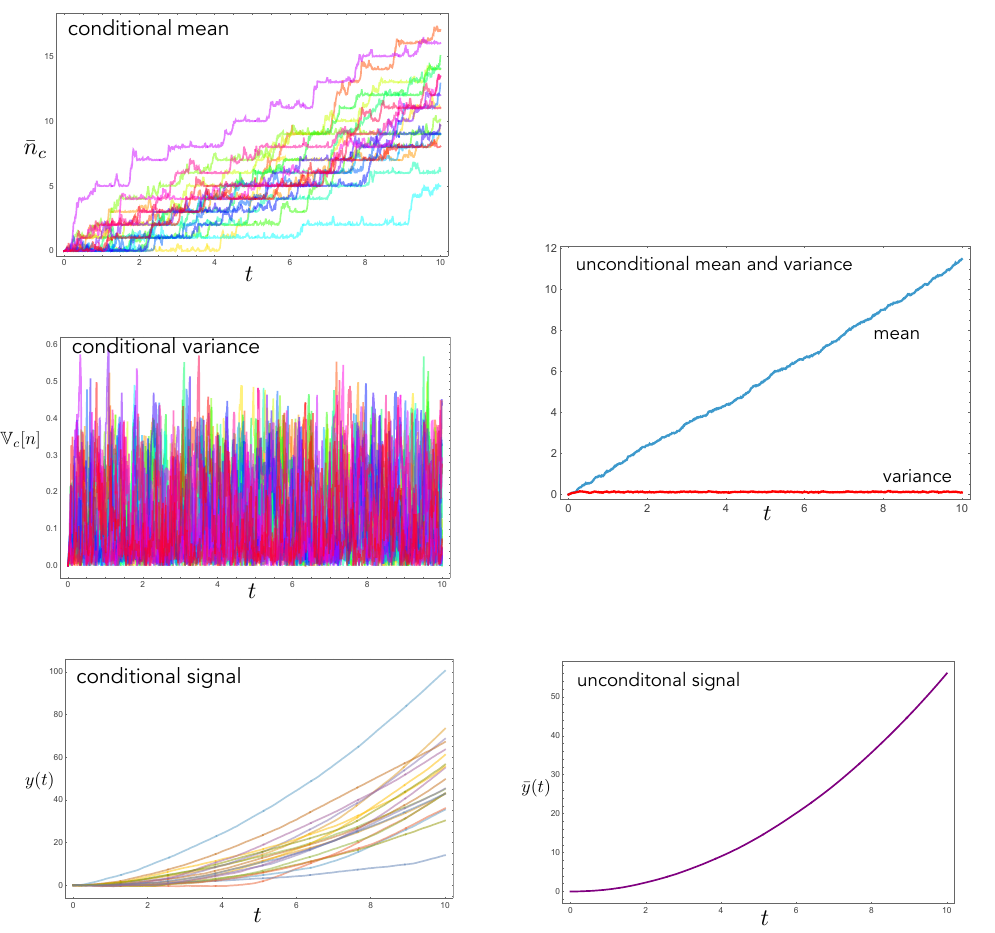}
    \caption{Strong measurement 'jump' regime . Typical simulations of the conditional and unconditional mean population, conditional variance,  and the stochastic measurement signal as a function of time. Also shown is the ensemble average stochastic measurement signal $\bar{y}(t)$. Parameters are $\gamma= 1.0, \Gamma =50.0$.  }
    \label{no-feedback-strong}
\end{figure}

In Figs. (\ref{no-feedback-weak}, \ref{no-feedback-strong}) we give examples of the dynamics of the conditional and unconditional moments for different values of $\Gamma$. In Fig. (\ref{no-feedback-weak}), the measurement strength is small $\Gamma=0.1$. In the second case, Fig. (\ref{no-feedback-strong}), the measurement strength is large $\Gamma=50.0$.  There is a qualitative difference in the behaviour. When the measurement strength is small the population is not closely monitored and the value grows diffusively.  When the measurement strength is large the population is closely followed and the current exhibits 'jumps'. This is similar to quantum jumps in the stochastic Schr\"{o}dinger equation\cite{WM}. 
In the case of a strong measurement, the conditional statistics immediately deviates from Poissonian.   

The conditional moments cannot be observed directly. In Figs. (\ref{no-feedback-weak}, \ref{no-feedback-strong}) we also plot the signal $y(t)$   defined by
\begin{equation}
\label{int-current}
    y(t) =\int_0^t I(t') dt'
\end{equation}
This is the empirical quantity of interest and it does not exhibit the large fluctuations evident in the conditional moments. It is a stochastic variable and may be interpreted as the area under the instantaneous measurement current $I(t)$. 

\subsection{Feedback.}
We implement a feedback process by giving agents access to the measurement current transmitted to the entire population. The rate of joining the population is proportional to the instantaneous value of the stochastic measurement current, $I(t)$. In biological interpretation, this is the rate of production. 

The feedback strength can be positive or negative.  If it is positive, individuals join the space at a faster rate than $\gamma$, while if it is  negative, they defect and leave the population\cite{Akt2016}.  The joining/defection rate is given by
\begin{equation}
    \gamma\rightarrow \gamma(t) = \gamma(1+\lambda I(t))\ ,
\end{equation} where $\lambda$ has units of time.  If $\lambda <0$, individuals defect. If $\lambda >0$ individuals join. The corresponding conditional Markov process is now 
\begin{equation}
\label{SME-feedback}
   dp_{c,n}= \gamma (p_{c,n-1}-p_{c,n})dt +r\bar{n}_c (p_{c,n-1}-p_{c,n})dt +\frac{r}{\sqrt{\Gamma}}(p_{c,n-1}-p_{c,n})dW  +\sqrt{\Gamma}(n-\bar{n}_c)p_{c,n}dW(t)
\end{equation}
where $ r = \gamma  \lambda \chi $ and $\Gamma=\chi^2/\kappa$. The conditional mean number of individuals joining the population now follows
  \begin{equation}
  \label{full-conditional}
    d\bar{n}_c(t) =\gamma dt+ r\bar{n}_c(t) dt +\left [\frac{r}{\sqrt{\Gamma}}+\sqrt{\Gamma}{\mathbb V}_c[n]\right ]dW
\end{equation}
 The derivation of Eq.(\ref{conditional-mean})  assumes that  $\gamma/\kappa \ll 1$ while $\Gamma/\kappa \sim 1$ (see Appendix). We then see that $r/\gamma=  \kappa\lambda\sqrt{\Gamma/\kappa}\sim \kappa\lambda $.

If we average over the noise, interpreted as having no knowledge of the actual measurement signal at any time,  the unconditional mean population satisfies
 \begin{equation}
    d\bar{n}(t) = (\gamma+r\bar{n}(t)) dt 
\end{equation}
where we have used $\langle dW\rangle =0$. It corresponds to exponential growth ($r >0$) or exponential decline ($r<0$). In the latter case, there is a steady state given by $\bar{n}^{\infty} = \gamma/|r|>0$.  

The stochastic differential equation for the conditional mean, $\bar{n}_c(t)$, depends on the conditional variance, ${\mathbb V}_c[n]$.
As previously, if $\gamma$ is large enough, we can approximate the conditional variance is proportional to the conditional mean. In that case
  \begin{equation}
    d\bar{n}_c(t) =\gamma dt+ r\bar{n}_c(t) dt +\sqrt{\Gamma}\bar{n}_c(t)dW
\end{equation}
This equation corresponds to geometric Brownian motion (GBM) for the conditional mean, with a constant background rate.     In the limit $r>>\gamma$,  the growth of the population is dominated by access to the measurement signal ( quality of the information in the global signal is high),
  \begin{equation}
  \label{conditional-pop-poisson}
    d\bar{n}_c(t) = r\bar{n}_c(t) dt +\sqrt{\Gamma}\bar{n}_c(t)dW
\end{equation}

\begin{figure}
    \centering
    \includegraphics[scale=1.0]{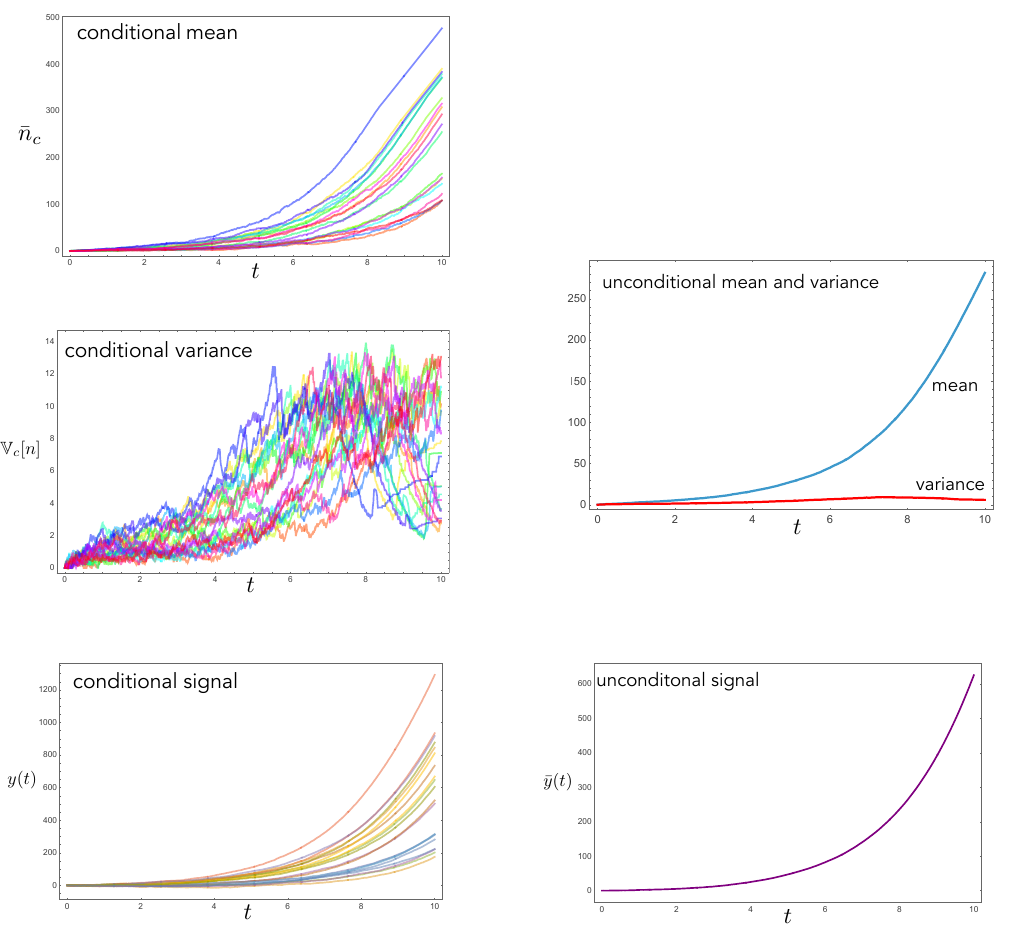}
    \caption {Sample trajectories for the conditional moments with feedback in the weak measurement regime.  The parameters are $\gamma=0.01, r=0.5, \Gamma =0.1$.   The conditional  measurement signal,$y(t)$, or accumulated wealth, now grows exponentially. }
    \label{feedback-weak}
\end{figure}
\begin{figure}
    \centering
    \includegraphics[scale=1.0]{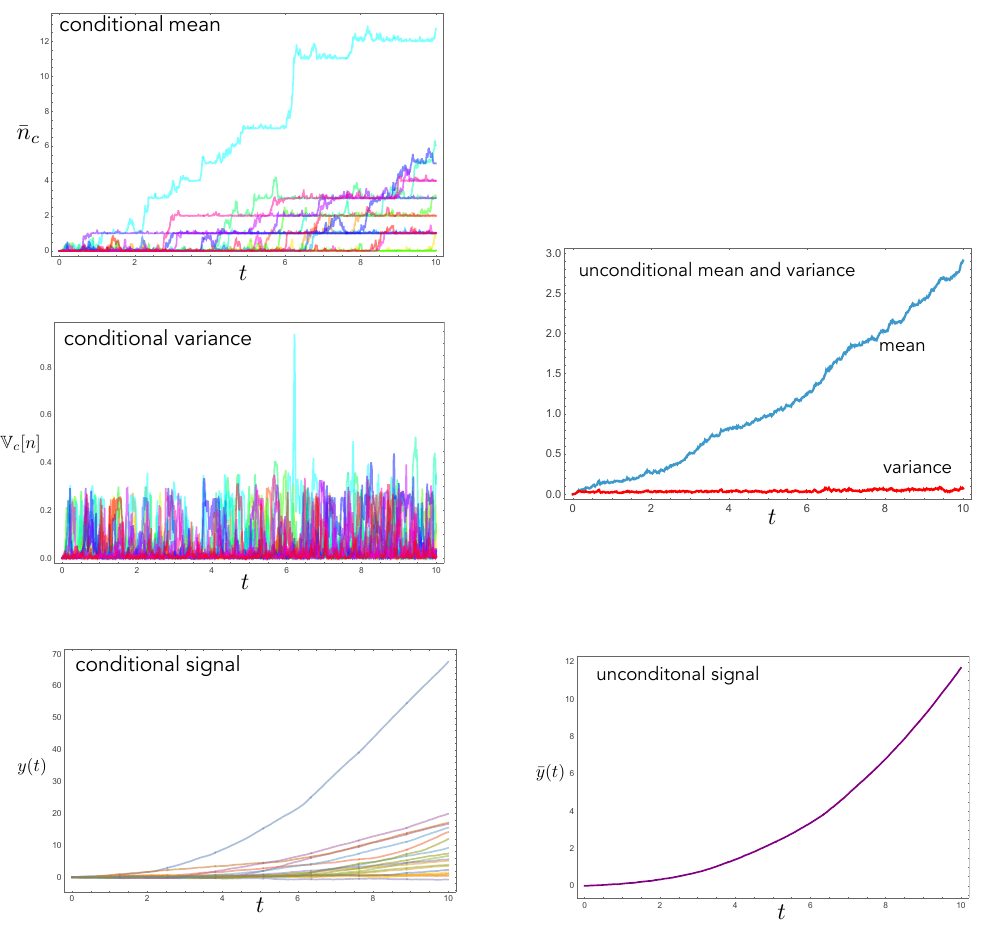}
    \caption {Sample trajectories for the conditional moments with feedback in the strong measurement regime, and  slow feedback. The parameters are $\gamma=0.1, r=0.1, \Gamma =50.0$.   }
    \label{feedback-strong1}
\end{figure}
\begin{figure}
    \centering
    \includegraphics[scale=1.0]{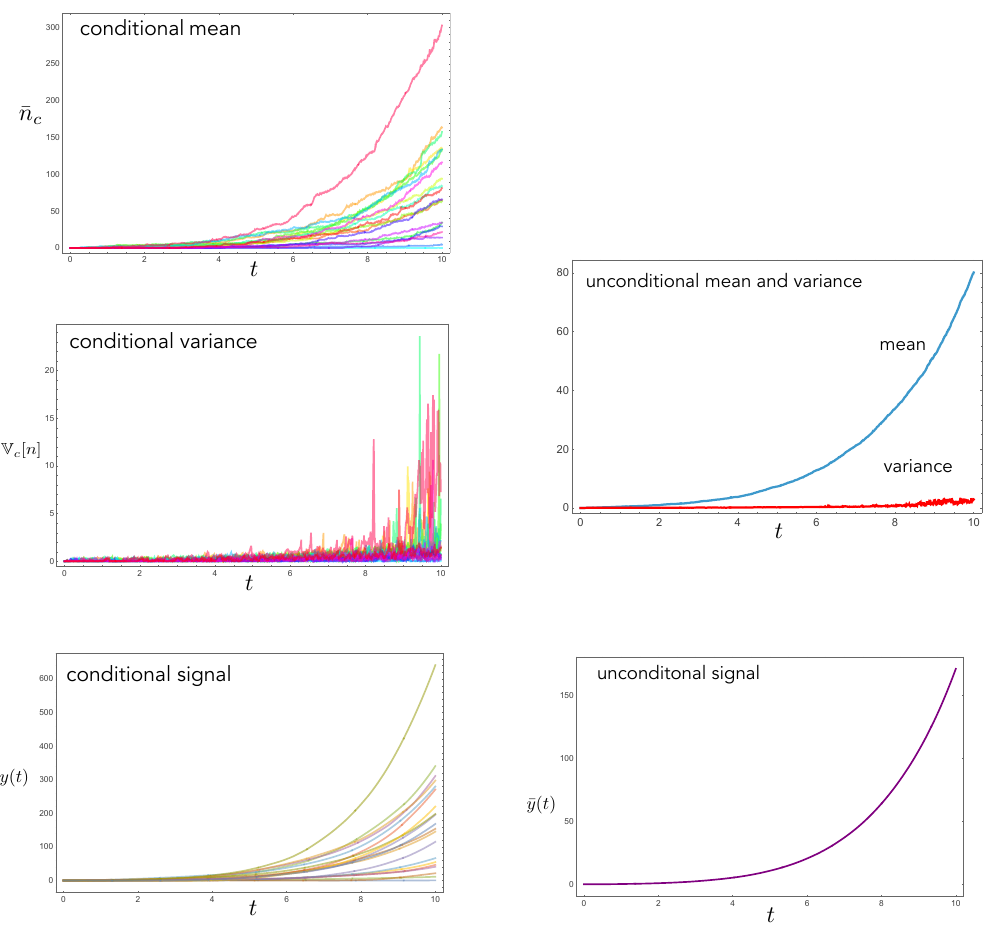}
    \caption {Sample trajectories for the conditional and unconditional  moments with feedback in the strong measurement regime and fast feedback. The parameters are $\gamma=0.1, r=0.5, \Gamma =50.0$.   The conditional  measurement signal, $y(t)$ grows exponentially.  }
    \label{feedback-strong2}
\end{figure}

In Fig. (\ref{feedback-weak},\ref{feedback-strong1},\ref{feedback-strong2})) we plot some sample trajectories for the conditional and unconditional moments with feedback. If  choose the feedback rate to be similar to the bare rate $r\sim \gamma$, the fluctuations dominate in both the weak and strong measurement regime. On the other hand, if the feedback rate is larger than the bare rate, the jump regime leads to a rapid 'collapse' of the population distribution to a single peak, Fig. (\ref{feedback-strong2}). In this case, the variance becomes small, and the exponential growth rate is determined by a single number.  This limit results in an average measurement signal that grows exponentially, so care must be taken with the truncation of the population in numerical simulations.  

\subsection{Comparison with geometric Brownian motion. }
Geometric Brownian motion is the central element in the Black-Scholes financial pricing model\cite{Black-Scholes}.  It has also been used as an essential element in theories of cooperation in populations\cite{Peters}. We discuss these more fully in the next section. Here we compare the stochastic dynamics of the feedback model we have proposed to GBM.

Geometric Brownian motion is defined by the Ito stochastic process,
\begin{equation}
    dS(t) =\mu S(t)dt +\sigma S(t) dW
\end{equation}
where the noise is multiplicative distinguishing the process from regular Brownian noise.  If we average over the noise, the mean satisfies the exponential growth equation
\begin{equation}
    \frac{d\bar{S}}{dt} =\mu \bar{S} 
\end{equation}
which would also be satisfied for the regular Brownian process $dS=\mu S(t)dt +\sigma dW$. 

Our model of a global measurement with feedback to population growth leads to GBM for the conditional average population $\bar{n}_c(t)$ {\em under the assumption that the conditional population statistics is a Poisson distribution.}  First define the stock price, $S(t)$ as the integral of the measurement current, 
\begin{equation}
    S(t)=\int_0^t I(t')dt'
\end{equation}
The stock price is then the measurement signal, $y(t)$. 
Using Eq. (\ref{measurement-current}), 
\begin{equation}
   dS = \chi \bar{n}_cdt+\sqrt{\kappa} dW
\end{equation}
where $\bar{n}_c$ is the predicted population, conditioned on previous stock prices and we interpret $\kappa$ as the volatility. When the background rate is sufficiently large, giving the approximation ${\mathbb V}_c[n]\approx \bar{n}_c$, the stochastic process $\bar{n}_c$ satisfies the GBM stochastic differential equation, Eq(\ref{conditional-pop-poisson}). 

Each individual acts completely independently but they all have equal access to the global quantity $S(t)$, the measurement signal in our description. This is the `free markets hypothesis'. Each individual is able to use this information to estimate the conditional mean number of agents $\bar{n}_c(t)$. At no point does anyone know the actual number of individuals in the market.

\section{Comparison to Peters' and Adamou's model of cooperation. }
 In \cite{Peters} Peters and Adamou (PA) address the question of why individuals should share resources if one is disadvantaged by the interaction. They propose a solution in terms of a stochastic growth model that distinguishes two kinds of average growth rates:  the ensemble-average growth rate, and the time-average growth rate. The latter is
smaller than the former by a term which depends on a diffusion rate in a multiplicative stochastic process. Their answer to the question turns on this difference.

The stochastic process considered in  \cite{Peters} is defined by the Ito stochastic differential equation for a stochastic real variables $x(t)$,
\begin{equation}
\label{GBM}
dx=x(t)(\mu dt+\sigma dW(t))
\end{equation}
where $ \mu, \sigma$ are real numbers, called the drift and the volatility respectively, and $dW(t)$ is the Wiener increment. This stochastic process is called Geometric Brownian Motion (GBM).  If the units are chosen so that $[t]=\mbox{s}$, then $[\mu]=\mbox{s}^{-1}, [\sigma]=\mbox{s}^{-1/2}$.  Typically $\mu >0$, in which case it is obvious that the stochastic process does not have a stationary state. 

We make a change of variable 
\begin{equation}
\label{change-variable}
s(t) =\ln x(t)
\end{equation}
Using the Ito calculus,  the stochastic differential equation for $s$ is
\begin{equation}
ds(t) = (\mu-\frac{\sigma^2}{2})dt  +\sigma dW
\end{equation}
The solution is 
\begin{equation}
\label{s-solution}
s(t) = s(0)+ (\mu-\frac{\sigma^2}{2})t+\sigma\int_0^t dW(t')
\end{equation}
Using the rules of stochastic integration, the ensemble mean and variance at time $t=T$ are then 
\begin{eqnarray}
{\mathbb E}[s(T)] & = & {\mathbb E}[s_0] +(\mu-\frac{\sigma^2}{2})T\\
{\mathbb V}[s(T)] & = & \sigma^2 T
\end{eqnarray}
where the averaging is done over infinitely many realisations of the stochastic trajectories, each starting with a value $s_0$ sampled from the initial distribution for $s$.  This is equivalent to Eq (2.2) in \cite{Peters}. We call this the `ensemble average' .Without loss of generality,  we will assume ${\mathbb E}[s_0]=0$. 

The change of variable in Eq.(\ref{change-variable}) ensures that $x(t)$ is always positive. If we were trying to find a diffusive model for a stochastic variable that is required to always be positive, this is one way to do it. In the case of the Black-Scholes pricing model, this is indeed a requirement as $x$ is a stock price. It is for this reason that Black-Scholes uses GBM. But it is not unique. 

The discussion in\cite{Peters} makes a distinction between time averages and ensemble averages for GBM,  Eq. (\ref{GBM}).   An ensemble is a theoretical construct as it is defined in term of an infinite sum over realisations.  In the ensemble average we assume that each realisation is independent so, by assumption, each element of the population is unaware of any other. There is no reciprocal interaction between the individuals. In our approach individuals only `see' others via the feedback of a global collective measurement current and that leads  to GBM for a given range of parameters.

Suppose now we are monitoring a particular stochastic trajectory of  $s(t)$ by making an arbitrarily accurate measurement of its value at times $t_n=n\tau$ for fixed $\tau$. This produces a time series  $s_1,s_2,\ldots s_n$.  We can then construct the empirical time average 
\begin{equation}
\bar{s}_N=\frac{1}{N}\sum_{k=1}^N s_k
\end{equation}
This is an estimate of the stochastic integral 
\begin{equation}
\label{stochastic-int}
\bar{s}(T)=\frac{1}{T}\int_0^T dt\  s(t)
\end{equation}
where $T=N\tau$. Note that $\bar{s}(T)$ is a random variable: it is the area under the stochastic trajectory $s(t)$ between $t=0$ and $t=T$, divided by $T$.  If we use Eq. (\ref{s-solution}) we find that this integral diverges as $T\rightarrow \infty$. This is not surprising as it is not a stationary process. Clearly the stochastic integral in Eq. (\ref{stochastic-int}) does not converge to ${\mathbb E}[s(T)] $ and the process is not ergodic.

PA postulate that what appears as cooperation is in fact due to the stochastic growth of resources via GBM:  "This allows us to study cooperation under simple conditions, where the effects invoked in classical explanations do not exist."\cite{Peters}. They emphasise the different growth rates for the ensemble averaged growth rate, $\mu$ of $x(t)$ versus the time average growth rate $\mu-\sigma^2/2$ of the derived stochastic variable $s(t)$ that follows from the Ito correction.  This makes it appear like a mathematical artefact. However, this is justified by the fact that $x(t)$ must remain positive to make physical sense. PA interpret this to mean\cite{Peters}
\begin{quote}
...repeated pooling and sharing reduces the net effect of
fluctuations, thereby increasing the time-average growth rate of each cooperator’s resources, which approaches the ensemble-average growth rate as the number of cooperators increases. Therefore, cooperation in our model is advantageous for the simple reason that those who do it outgrow those who do not.
\end{quote}
 A minimal postulate consistent with PA's claims is that populations of individuals whose stochastic growth follows GBM will out-grow those populations that follow only constant diffusion with exponential growth. 

Our proposal, based on feedback of a global measurement current,  provides an explicit model for a wide class of  growth models for positive-valued stochastic processes, $x(t)>0$, including GBM in an appropriate limit. The global measurement current available equally to all individuals in a population is a kind of sharing or pooling. In a financial setting it is an implementation of the efficient markets hypothesis.

 This results in dynamics similar to the PA model\cite{Peters} (when $r\gg\gamma$).  The conditional equation, however has a different interpretation to that given by PA. It is a conditional average conditioned on the entire history of the measurement current, $I(t)$. If a mean is conditioned on the past history it cannot be ergodic.

\section{Sacrifice: two phenotypes. }

In \cite{Ackermann} a model for self-destructive cooperation mediated by phenotypic noise is presented. A population of genetically identical organisms, living in the same environment, expressing two kinds of phenotypes is called phenotypic noise. It refers to the variability in observable traits (phenotypes) among individuals that share the same genotype and are in the same environment.  The two phenotypes are labelled  cooperators (C) and defectors (D). The cooperators self-destruct and contribute a public good to the population. The defectors survive and form a successful lineage all expressing the same phenotype.

Let $p^\alpha_n(t)$, where $\alpha\in \{C,D\}$, be the probability that there are $n$ individuals of phenotype $\alpha$ in the space at time $t$. The probabilities satisfy the Markov process,
\begin{eqnarray}
\frac{dp^D_n}{dt}& = & \gamma (p^D_{n-1}-p^D_n) \\
\frac{dp^C_n}{dt} & = &  \gamma (p^C_{n-1}-p^C_n) +\kappa (n+1) p^C_{n+1} -\kappa np^C_n
\end{eqnarray}
where $\gamma$ is the rate at which individuals of both types enter the space while cooperators die  at a rate $\kappa$. 
Assuming $p^D_n(0)=\delta_{n,0}$ the solution for defectors is 
\begin{equation}
    p^D_n(t) = \frac{(\gamma t)^n}{n!} e^{-\gamma t}
\end{equation}
The mean number and variance  are ${\mathbb V}[n^D]=\bar{n}^D(t)=\gamma t$.  Assuming $p^C_n(0)=\delta_{n,0}$, the solution for cooperators is
\begin{equation}
    p^C_n(t)=\exp\left [-\frac{\gamma}{\kappa}(1-e^{-\kappa t})\right ]\left (\frac{\gamma}{\kappa}\right )^{n}(1-e^{-\kappa t})^n
\end{equation}
The cooperators have a steady state solution is a Poisson distribution, 
\begin{equation}
    p^C_n(\infty)= \frac{1}{n!}\left (\frac{\gamma}{\kappa}\right )^{n} e^{-\frac{\gamma}{\kappa}}
\end{equation}
The mean number of cooperators with this initial condition is
\begin{equation}
    \bar{n}_C(t) = \frac{\gamma}{\kappa}\left ( 1-e^{-\kappa t}\right )
\end{equation}
The probability for the population to have $n$ individuals without regard for phenotype is 
\begin{equation}
    p_n(t)=p^C_n(t)+p^D_n(t)
\end{equation}
Clearly the total mean population size is 
\begin{equation}
    \bar{n}(t) = \gamma t + \frac{\gamma}{\kappa}\left ( 1-e^{-\kappa t}\right ) 
\end{equation}
In the limit $\kappa t>>1$, the total population is given by 
\begin{equation}
    \bar{n}(t)\approx \gamma t+\frac{\gamma}{\kappa}\ \ \ \mbox{for  }\ \  \kappa t >>1
    \end{equation}
There is no steady state for defectors, as in the single phenotype  model.

We assume that the measurement signal is simply proportional to the {\em total }population given by the value of the stochastic process $n(t)= n^C(t)+n^D(t)$ at any time. The measurement signal is subject to diffusive (Wiener)fluctuations with constant variance, as in Eq.(\ref{measurement-current})
\begin{equation}
   I(t)dt= \chi \bar{n}_c(t) dt+\sqrt{\sigma}dW(t)
\end{equation}
where $dW$ is a Wiener increment,  $\bar{n}_c(t)$ is the conditional mean total population conditioned on the entire stochastic history of $I(t)$ up to time $t$, $\chi$ is a rate that determines the instantaneous value of the measurement signal and $\sigma$ sets the size of the diffusive noise. 
The conditional mean thus obeys the stochastic differential equation, Eq. (\ref{SME}).   
 
As an example of cooperator behaviour we offer bacillus subtilis populations. In this case a sub population of cells, `co operators', produce  proteases that degrade
proteins into smaller peptides. These are reabsorbed as a nutrient source by all cells, including the subpopulation of cells that do not sacrifice themselves that we call `defectors'\cite{Cooper}. 

This leads us to define another measurement current, the benefit current.  In effect, this additional measurement process continuously monitors the number of cooperators that decay from the population as a `mortality' counter $M(t)$. This is a Poisson process (jump process) with stochastic increments, $dM(t)\in \{0,1\}$, with 
\begin{equation}
{\mathbb E}[dM(t)] =\kappa\bar{n}^C(t)dt
\end{equation}
and where $\bar{n}^C(t)=\sum_n np^C_n(t)$.  The benefit measurement signal is then defined by the Ito stochastic differential equation 
\begin{equation}
   I_b(t)dt= \bar{n}^C_c(t) dt+\sqrt{\rho}dW_C(t)
\end{equation}
where $dW_C$ is a Wiener increment,  $\bar{n}^C_c(t)$ is the conditional mean cooperator population conditioned on the stochastic history of $I_b(t)$ up to time $t$, $\rho$ sets the size of the diffusive measurement noise in the monitoring of the cooperator population.  The corresponding stochastic differential for the conditional mean $\bar{n}^C_c(t)$ is
\begin{equation}
   d\bar{n}^C_c =\gamma dt-\kappa \bar{n}^C_c dt+\sqrt{\Gamma_C}{\mathbb V}_c[n^C]dW^C 
\end{equation}
where $\Gamma_C$ is the measurement rate for observation of the death of the cooperators through partial observations of the point process $M(t)$.
Knowledge of both $I(t)$ and $I_b(t)$ fixes the conditional state of defectors as 
\begin{equation}
    I_D(t) dt= (I(t)-\chi I_b(t))dt= \chi \bar{n}^D_c dt+dW_D(t)
\end{equation}
\begin{equation}
   dW_D(t) = \sqrt{\sigma}dW(t)-\chi\sqrt{\rho}W_C
\end{equation}
We now have two global measurements, one responding to the total population and one responding only to the the population of cooperators.

Although both phenotypes contribute to the measurement current equally, the response depends on the phenotype.  Now we implement a feedback process by giving agents access to the measurement signals, that is, they can access the information contained in the environment by making the rate of joining the space conditional on the instantaneous value of the stochastic  measurement signals, $I(t), I_b(t)$. Feedback can be positive  or negative.  If it is positive,  individuals join the space at a faster rate than otherwise, while if it is negative,  individuals defect and leave the population\cite{Akt2016}.   

\begin{figure}
      \centering
      \includegraphics[scale=0.75]{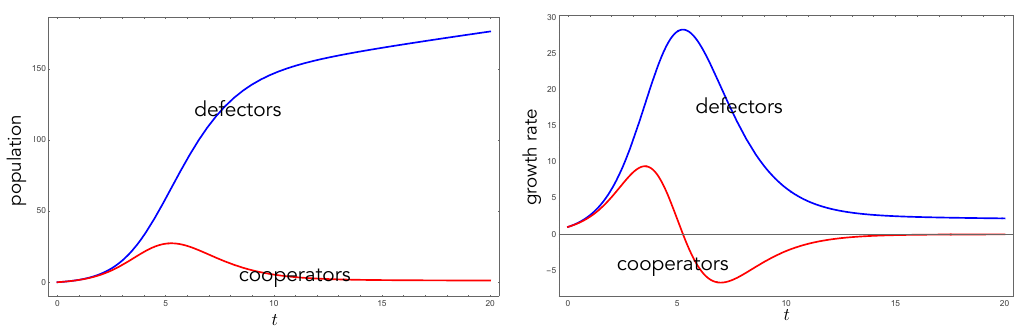}
      \caption{Left: The average number of cooperators and defectors as a function of time. Right: average growth rates.   $\kappa=0.1, \nu=0.1, \lambda=1.0$. }
      \label{coperate-rate}
  \end{figure}

We let the joining rate for defectors and collaborators joining the space depend on the benefit current by $\gamma\rightarrow \gamma(t) = \gamma(1+\lambda I_b(t))$, where $\lambda > 0$.  We let the death rate of cooperators depend on the total population current, $I(t)$ as $\kappa\rightarrow \kappa(1+\nu I(t))$  The conditional mean number of individuals of each type joining each population now follows 
  \begin{eqnarray}
      d\bar{n}^D_c(t) & = & \gamma(1+\lambda I_b(t))dt+\sqrt{\Gamma}{\mathbb V}_c[n^D]dW\\
       d\bar{n}^C_c(t) & = &  \gamma(1+\lambda I_b(t))dt-\kappa(1+\nu I(t))\bar{n}_C dt+\sqrt{\Gamma_C}{\mathbb V}[n^C]dW_C
  \end{eqnarray} 
  The independent Weiner increment $dW_C$ arises from partially observed process that monitors the population of cooperators. 
If we average over the noise, interpreted as having no knowledge of the actual stochastic signal at any time,  the unconditional mean population obey the non linear system 
  \begin{eqnarray}
      \frac{d\bar{n}^D}{dt} & = & \gamma(1+\lambda  \bar{n}^C)\\
       \frac{d\bar{n}^C}{dt} & = &  \gamma(1+\lambda  \bar{n}^C)-\kappa(1+\nu  (\bar{n}^D+\bar{n}^C))\bar{n}^C 
  \end{eqnarray}
  Here $\kappa$ is the bare death rate for cooperators, $\nu$ is the feedback enhancement of the death rate, $\lambda$ is the benefit feedback enhancement of the joining rate.
  It is convenient to use units of time such that $\gamma=1$ and $\kappa\rightarrow \kappa/\gamma$).  In Fig.(\ref{coperate-rate}) we plot some examples.  

\section{Discussion.}
Geometric Brownian motion is a widely used stochastic model for multiplicative noise and is commonly applied in settings where a stochastic variable must remain positive, such as population size or asset prices. In this paper, we have shown that GBM arises as a special case of a more general class of dynamics generated by a partially observed continuous-time Markov process subject to feedback from a global measurement signal. In this framework, individuals do not observe the macroscopic state directly. Instead, they condition their behaviour on a shared but noisy signal that is correlated with an underlying collective variable, while the true population state remains hidden. Averaging over the observed process yields exponential growth, while individual stochastic trajectories may be diffusive or jump-like. Which regime dominates depends on the signal-to-noise properties of the measurement channel, even though the ensemble-average trajectory remains exponential.

By construction, this approach avoids assumptions that underpin many standard models of coordination and cooperation. No agent optimises an objective function, implements a learning algorithm, or communicates directly with others. Collective behaviour instead arises from macro-to-micro feedback mediated by partial observability. From this perspective, behaviours often described as cooperation can be interpreted more precisely as coordination: alignment of individual dynamics induced by shared access to imperfect information about an emergent macroscopic state. The distinction between ensemble-average and time-average growth rates emphasised in stochastic growth models appears here as a property of conditional dynamics under measurement, rather than as evidence for cooperative intent or strategic interaction.

The framework applies to biological and economic systems in which collective outcomes depend on indirect, noisy coupling between local actions and aggregate states. In biological contexts, the measurement signal may represent a metabolic or ecological output that correlates with population activity and modulates growth. In financial markets, it may be interpreted as a market signal through which individuals condition their behaviour on aggregate information without direct access to underlying state variables. In both cases, abrupt transitions, volatility, or apparent coordination failures arise not from antagonism or defection, but from limitations in the information channel linking micro-level dynamics to macroscopic outcomes.

Extending the model to populations with multiple phenotypes illustrates how sacrifice-like behaviour can be favoured when rapid early growth confers an advantage. Here, feedback between phenotypic composition and global signals gives rise to dynamics resembling autocatalytic chemical reactions. These results show that cooperation-like outcomes can arise within a purely dynamical, physically grounded framework, without assuming altruism, norms, or explicit strategic reasoning.

The present framework also suggests a complementary way of interpreting global collective action problems such as climate change, which are commonly diagnosed in mainstream economics as market failures arising from unpriced externalities. From the viewpoint adopted here, prices, taxes, standards, and other policy instruments can be understood as measurement signals intended to proxy emergent variables—such as the social cost of carbon—that arise from the coupled dynamics of climate and society and are not directly observable. Fully internalising externalities would therefore require these signals to reliably encode highly complex, uncertain, and evolving macroscopic states, an informational requirement that is unlikely to be achievable in practice. Within this perspective, the effectiveness of coordination mechanisms depends less on whether an externality is “fully priced” in principle than on the signal-to-noise properties of the feedback channels through which agents condition their behaviour.

More broadly, the model invites a reconsideration of how preferences are represented in theories of collective behaviour. In many standard approaches, subjective utilities or payoff functions are treated as primitives external to the model, rather than as consequences of physically implementable mechanisms. In the present framework, the analogue of a preference is the feedback law by which a partially observed measurement signal modulates microscopic transition rates. Collective behaviour can thus be described as inference and control on hidden macroscopic states, with coordination performance constrained by the reliability of the information channel rather than by assumed objectives. The emergence of distinct dynamical regimes as the signal-to-noise ratio varies points towards a more fully physical account of collective behaviour, connecting coordination capacity to the thermodynamics of measurement, computation, and feedback.

Several directions for future investigation follow naturally. Here we have assumed a simple, instantaneous form of feedback from an exogenously defined global measurement signal. More general models could allow the measurement signal itself to exhibit endogenous stochastic dynamics, including nonlinear interactions and feedback between the signal and the population it monitors. Such feedback-controlled measurement processes arise naturally in quantum measurement and control \cite{PhysRevLett.129.050401}, where information, dynamics, and control are treated on equal physical footing. In economic contexts, and financial markets in particular, endogenous modification or control of measurement signals could significantly alter individual and collective dynamics. The adaptive markets hypothesis \cite{AdaptiveMarketsHypothesis} suggests many examples in which agents actively shape the informational environment to which others respond. While overt manipulation, such as insider trading, is often detectable, more subtle, high-frequency, or distributed modifications of informational cues may be harder to identify—especially in regimes dominated by jump-like processes, such as those observed in cryptocurrency markets. Extending the present framework to include endogenous and strategically modified signals would make it possible to study how such interventions affect coordination under partial observability. In this setting, machine learning techniques may provide useful tools for detecting hidden structure or dynamical regularities in observed signals and for inferring underlying feedback processes that are not directly accessible. Taken together, these extensions point towards a physically grounded theory of collective behaviour in which coordination, cooperation and their failures emerge from the dynamics of information, measurement, and feedback under fundamental epistemic constraints.  

\section*{Appendix}
\begin{figure}[h!]
\centering
\includegraphics[scale=0.5]{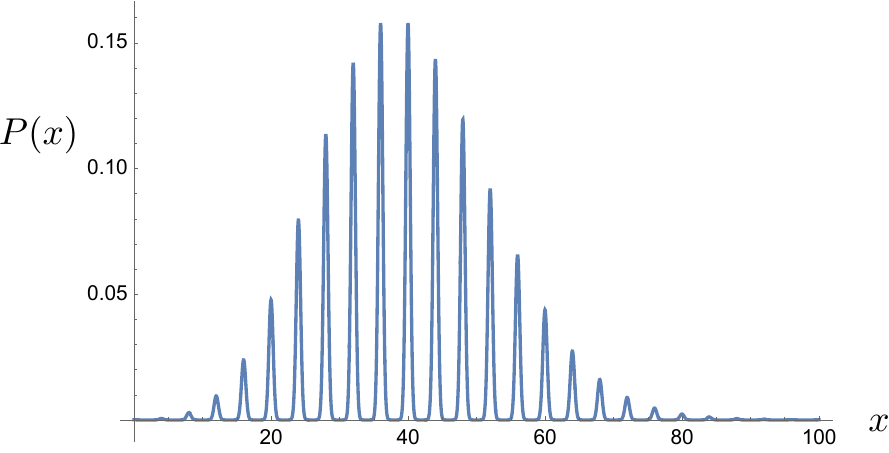}
\caption{A plot of the observed distribution $P(x)$, Eq.(\ref{obs})}, assuming the underlying population distribution is a Poisson distribution with mean $10$. The other parameter values are $\Delta=0.1$ and $\mu=4$. 
\label{obs-distr}
\end{figure}

 The state of the population at any time is defined by the probability distribution $p_{n}(t)$. Our objective is to estimate this state by making imperfect measurements on the population. We begin by supposing that an indirect measurement is made of the number, $n$, by making an observation on a real valued quantity,  $x\in{\mathbb R}$ that is correlated with $n$.     We define the response of this variable to the population count, $n$, by a conditional probability $P(x|n)$. The unconditional statistics of the output variable $x$, is given by 
\begin{equation}
\label{obs}
    P(x,t)=\sum_{n=0}^\infty P(x|n)p_n(t).
\end{equation}
We may also calculate the \emph{conditional} state of the population given a particular observation $x$. This is given by Baye's rule as 
\begin{equation}
  p_n(t|x) = \frac{P(x|n)p_n(t)}{P(x,t)}.
\end{equation}
As an example, consider the Gaussian
\begin{equation}
    P(x|n) =(2\pi\Delta)^{-1/2} e^{-(x-\mu n)^2/2\Delta},
\end{equation}
where $\mu$ is a constant, with the same units as $x$, which defines the separation of Gaussian peaks corresponding to the different values of $n=\pm 1$ in the distribution of $P(x)$ , and $\Delta$ defines the width of each peak (see Fig. (\ref{obs-distr})). While $\mu$ depends on how strongly the population count is coupled to the observed quantity, $\Delta$ depends on the number of measurements we perform to sample $x$, hence, in the case of a very poor measurement, $\Delta >>\mu^2$. 

It is easy to see that the mean and variance of the observed  process  are given by
\begin{eqnarray}
\bar{x}  & = &  \mu\bar{n},\\
{\mathbb V}[x] &= & \Delta +4 \mu^2 {\mathbb V}[n].
\end{eqnarray}
We see the noise is additive: the first term in ${\mathbb V}[x]$ shows the measurement noise and the second term is proportional to the variance of the population. 

We can define a diffusive continuous-time version of this measurement by making a sequence of very many such measurements repeated  $N>>1$ times in a time interval $[t, t+\delta t)$ such that $\gamma\delta t <<1$. That is to say, there are very many measurements made on a time scale short compared to the background system dynamics.  Let us now form the effective time average 
\begin{equation}
    y(t) =\frac{1}{N}\sum_{j=1}^N x_j.
\end{equation}
The law of large numbers then implies that the mean and variance of $y(t)$ are given by  
\begin{eqnarray}
\bar{y}(t) & = & \mu \bar{n},\\
{\cal V}[y(t)] & = & \frac{\Delta}{N}.
\end{eqnarray}
We introduce the rate of measurement as $\kappa=N/\delta t$ such that $\sqrt{\Delta}=  \kappa \delta t $, and further define $\mu =\chi \delta t$. The continuous limit is now taken by assuming that as $N=\kappa \delta t \rightarrow \infty $, the ratio $\mu/\sqrt{\Delta} =\chi/\kappa$ is fixed. This fixes the ratio of the peak separation to the peak width of the gaussian peaks in the distribution for $P(x)$ as the continuous limit is taken. Then, we see that  ${\cal V}[y(t)] \propto \delta t$. This is a diffusion process, so we can write
\begin{equation}
\label{signal-sp}
dy(t) = \chi \bar{n} dt+\sqrt{\kappa}dW.
 \end{equation}
If the system is in one state or the other at time $t$, the observed process is subject to diffusion at rate $\kappa$. Finally we write this in terms of a stochastic current using $dy(t) = I(t) dt$. 
\begin{equation}
I(t)dt = \chi \bar{n} dt+\sqrt{\kappa}dW.
 \end{equation}
 
In a similar way we can compute the conditional probability distribution of the population given a particular result $y$ in a time $\delta t$. Let $p_{c,n}$ be the conditional state of the process given an entire history of results $y(t)$ up to time $t$.  Using Baye's theorem we see that
\begin{equation}
\label{update}
p_{c,n|y}=\frac{P(y|n)p_{c,n}}{P(y)}.
\end{equation}
where 
\begin{equation}
    P(y|n)=(2\pi\Delta/N)^{-1/2}\exp\left [-\frac{(y-\chi n)^2}{2\Delta/N}\right ]
\end{equation}
and 
\begin{equation}
    P(y)=\sum_{n=0}^\infty p_nP(y|n)
\end{equation}
If we expand Eq. (\ref{update}) to linear order in $\chi$ and use Eq.(\ref{signal-sp}), taking care that $dW^2\sim dt$ we find that in the continuous time limit the conditional distribution satisfies the Ito stochastic differential equation 
\begin{equation}
   dp_{c,n}= \gamma (p_{c,n-1}-p_{c,n})dt   +\sqrt{\Gamma}(n-\bar{n}_c)p_{c,n}dW(t)
\end{equation}
where  $\Gamma =\chi^2/\kappa$ and
\begin{equation}
    \bar{n}_c(t)=\sum_{n=0}^\infty n p_{c,n}(t)
\end{equation}
is the mean conditioned on the entire history of the stochastic process $I(t)$. Recall that the continuous diffusive limit assumes that $\kappa >>\gamma$ but that the ratio $\chi/\kappa$ is fixed. This implies that $\Gamma/\kappa =\left (\chi/\kappa \right )^2$ is also fixed. This does not constrain the ratio $\Gamma/\gamma$. 


\bibliographystyle{alpha}
\bibliography{ringsmuth}

\newcommand{\etalchar}[1]{$^{#1}$}
\begin{thebibliography}{AABB{\etalchar{+}}22}

\bibitem[A.16]{Akt2016}
Aktipis A.
\newblock Principles of cooperation across systems: from human sharing to multicellularity and cancer.
\newblock {\em Evol. Appl.}, 9:17--36, 2016.

\bibitem[AABB{\etalchar{+}}22]{PhysRevLett.129.050401}
Bj\"orn Annby-Andersson, Faraj Bakhshinezhad, Debankur Bhattacharyya, Guilherme De~Sousa, Christopher Jarzynski, Peter Samuelsson, and Patrick~P. Potts.
\newblock Quantum fokker-planck master equation for continuous feedback control.
\newblock {\em Phys. Rev. Lett.}, 129:050401, Jul 2022.

\bibitem[ASF08]{Ackermann}
M.~Ackermann, B.~Stecher, and N.~et~al. Freed.
\newblock Self-destructive cooperation mediated by phenotypic noise.
\newblock {\em Nature}, 454:987–990, 2008.

\bibitem[BK13]{BurgerKliaras2013JumpDiffusion}
Patrick Burger and Marcus Kliaras.
\newblock Jump diffusion models for option pricing vs. the black scholes model.
\newblock Working Paper Series WP 081, University of Applied Sciences bfi Vienna, 2013.
\newblock Working Paper Series No. 81.

\bibitem[BS73]{Black-Scholes}
F.~Black and M.~Scholes.
\newblock The pricing of options and corporate liabilities.
\newblock {\em Journal of Political Economy}, 81:637--654, 1973.

\bibitem[BS20]{benjamin2020}
Colin Benjamin and Shubhayan Sarkar.
\newblock Emergence of cooperation in the thermodynamic limit, 2020.

\bibitem[CFL09]{RevModPhys.81.591}
Claudio Castellano, Santo Fortunato, and Vittorio Loreto.
\newblock Statistical physics of social dynamics.
\newblock {\em Rev. Mod. Phys.}, 81:591--646, May 2009.

\bibitem[CLP22]{Cooper}
G.A. Cooper, M.~Liu, and J.~et~al. Pena.
\newblock The evolution of mechanisms to produce phenotypic heterogeneity in microorganisms.
\newblock {\em Nat Commun}, 13:195, 2022.

\bibitem[DS04]{ACO}
Marco Dorigo and Thomas St\"{u}zle.
\newblock {\em Ant Colony Optimization}.
\newblock The MIT Press, 2004.

\bibitem[FMPG23]{PhysRevE.108.L012401}
Lorenzo Fant, Onofrio Mazzarisi, Emanuele Panizon, and Jacopo Grilli.
\newblock Stable cooperation emerges in stochastic multiplicative growth.
\newblock {\em Phys. Rev. E}, 108:L012401, Jul 2023.

\bibitem[Hau10]{Haugh2010BeyondBlackScholes}
Martin Haugh.
\newblock Beyond black-scholes.
\newblock Lecture notes for IEOR E4707: Financial Engineering: Continuous-Time Models, 2010.
\newblock Fall 2010.

\bibitem[Kau93]{Kauffman1993OriginsOfOrder}
Stuart~A. Kauffman.
\newblock {\em The Origins of Order: Self-Organization and Selection in Evolution}.
\newblock Oxford University Press, New York, 1993.

\bibitem[LZ24]{AdaptiveMarketsHypothesis}
Andrew~W. Lo and Ruixun Zhang.
\newblock {\em The Adaptive Markets Hypothesis: An Evolutionary Approach to Understanding Financial System Dynamics}.
\newblock Oxford University Press, Oxford, 2024.

\bibitem[OA22]{Peters}
Peters O and Adamou A.
\newblock The ergodicity solution of the cooperation puzzle.
\newblock {\em Phil. Trans. R. Soc. A}, 380:20200425, 2022.

\bibitem[RZ24]{RL-optimisation}
Tianyu Ren and Xiao-Jun Zeng.
\newblock Reputation-based interaction promotes cooperation with reinforcement learning.
\newblock {\em IEEE Transactions on Evolutionary Computation}, 28(4):1177--1188, 2024.

\bibitem[SP73]{Smith-Price}
J.~Maynard Smith and G.~R. Price.
\newblock The logic of animal conflict.
\newblock {\em Nature}, 246, 1973.

\bibitem[SSB{\etalchar{+}}20]{e22121432}
Viktor Stojkoski, Trifce Sandev, Lasko Basnarkov, Ljupco Kocarev, and Ralf Metzler.
\newblock Generalised geometric brownian motion: Theory and applications to option pricing.
\newblock {\em Entropy}, 22(12), 2020.

\bibitem[SUBK19]{PhysRevE.99.062312}
Viktor Stojkoski, Zoran Utkovski, Lasko Basnarkov, and Ljupco Kocarev.
\newblock Cooperation dynamics in networked geometric brownian motion.
\newblock {\em Phys. Rev. E}, 99:062312, Jun 2019.

\bibitem[VCBJ{\etalchar{+}}95]{vicsek1995novel}
Tam{\'a}s Vicsek, Andr{\'a}s Czir{\'o}k, Eshel Ben-Jacob, Inon Cohen, and Ofer Shochet.
\newblock Novel type of phase transition in a system of self-driven particles.
\newblock {\em Physical review letters}, 75(6):1226, 1995.

\bibitem[WM09]{WM}
H.M. Wiseman and G.J. Milburn.
\newblock {\em Quantum {Measurement} and {Control}}.
\newblock Cambridge University Press, 2009.

\end{thebibliography}

\end{document}